# The unmasking of 'Mitochondrial Adam' and Structural Variants larger than point mutations as stronger candidates for traits, disease phenotype and sex determination

**Abhishek Narain Singh**
**Web: ABioTek www.tinyurl.com/abinarain**
**abhishek.narain@iitdalumni.com**

**Abstract**

**Background**: Structural Variations, SVs, in a genome can be linked to a disease or characteristic phenotype. The variations come in many types and it is a challenge, not only determining the variations accurately, but also conducting the downstream statistical and analytical procedure.

**Method**: Structural variations, SVs, with size 1 base-pair to 1000s of base-pairs with their precise breakpoints and single-nucleotide polymorphisms, SNPs, were determined for members of a family. The genome was assembled using optimal metrics of ABySS and SOAPdenovo assembly tools using paired-end DNA sequence.

**Results**: An interesting discovery was the mitochondrial DNA could have paternal leakage of inheritance or that the mutations could be high from maternal inheritance. It is also discovered that the mitochondrial DNA is less prone to SVs rearrangements than SNPs, which propose better standards for determining ancestry and divergence between races and species over a long-time frame. Sex determination of an individual is found to be strongly confirmed using calls of nucleotide bases of SVs to the Y chromosome, more strongly determined than SNPs. We note that in general there is a larger variance (and thus the standard deviation) in the sum of SVs nucleotide compared to sum of SNPs of an individual when compared to reference sequence, and thus SVs serve as a stronger means to characterize an individual for a given trait or phenotype or to determine sex. The SVs and SNPs in HLA loci would also serve as a medical transformational method for determining the success of an organ transplant for a patient, and predisposition to diseases apriori. The sample anonymous dataset shows how the de-novo mutation can lead to non-inherited disease risk apart from those which are known to have a disease to mutation association. It is also observed that mtDNA is highly subjected to mutation and thus the factor for a lot of associated maternally inherited diseases.

**Conclusion**: 'mitochondrial Adam' can be a fair reality as certainly the biparental mode of mtDNA puts in question the theory of 'mitochondrial Eve'. SVs would serve as a stronger fingerprint of an individual contributing to his traits, sex determination, and drug responses than SNPs.

## 1 Background

Customization of the genome and biological material analysis for a tailor-made solution individual specific is the next bio and medical selling proposition. The high false discovery rate of structural variation algorithms, even in deeply sequenced individual genomes of the order of 30x average coverage [1, 2], suggests that for lower coverage the larger problem will to get rid of false positives. Nevertheless, the results with coverage as low as 3-5x also could have meaningful findings and be deployed for several genomes analysis, which would make sense on a population-wide scale at a relatively lower cost, such as in the 1000 genomes project, phase I [36].

In the current article, we discuss and extend the findings of mtDNA and Y-chromosome characteristics in terms of SNPs and SVs which was first presented in the year 2012 at a personalized genomics conference [23], followed by an extended article in 2013 at the international BIOCOMP conference [24], and it would be highly unlikely that many such similar experiments would not converge to some discoveries as it did happen in later years [25, 26, 27]. This paper serves as an extended paper of [23,24,49,50], particularly adding to them the clinical procedure following the variants extraction, and emphasis and clarity on mitochondrial inheritance possibility through paternal lines and thus the associated diseases. Most recently, Luo et. al. [46] in 2018, also came up with a large cohort that showed it is possible to have mtDNA being biparentally inherited, a discovery which was already made 6 years ago by article [23] in the year 2012, although a rare case of paternal mtDNA inheritance was also reported earlier [47]. The article thus has now been appropriately titled as mitochondrial Adam, to give due weightage of possibility of paternal or bi-paternal inheritace of mitochondrial DNA as previously mitochondrial Eve [48] paper discussed the possibility of maternal inheritance of mitochondria as the only possibility. In addition, the paper also highlights the rising importance of mutations larger than just the point mutations, as a significant contributing factor to determine end phenotype and trait.

With regards to discovery of SVs association to traits and phenotype of individual, here, more precisely we examine the InDels in SVs which are also known as DIPs. The analysis is then done for the whole family for matches of the SNPs variations to known pathogenic SNPs database, ClinVar [31, 37] for the sake of completion and way forward for medical diagnostics and was also presented at the international BIOCOMP conference [45] in July 2019.

Variations at specific loci in the genome have been associated with recurrent genomic rearrangements as well as with a variety of diseases, including color blindness, psoriasis, HIV susceptibility, Crohn's disease and lupus glomerulonephritis [3-8]. Figure F summarizes in a broad sense the various variations that can occur in a genome in comparison to a reference genome, as was initially published at [40]. This only enhances the importance of a comprehensive catalog for genotype and phenotype correlation studies [1-8] when the rare or multiple variations in gene underlie characteristic or disease susceptibility [9, 10]. Microarrays [11-13] and sequencing [14-17] reveal that the structural variants (SVs) contribution is significant in characterizing populations [18] and disease [19] characteristics. Interestingly the HLA domain in chromosome VI of an individual, which is the MHC region in humans, would be interesting in being decoded for the variations, as a lesser difference between two individuals could imply a greater likely of success in an organ transplant. With time the sequencing of human genomes now become routine [1], the spectrum of structural variants and copy number variants (CNVs) has widened to include much smaller events. The important aspect now is to know how genomes vary at large as well as fine scales and by what magnitude does it impact a population in general and an individual. There are databases such as OMIM [30], ClinVar[31, 37], dbSNP[32], PharmGKB[33], HGVbaseG2P [34] now named GWAS Central, UniProt [35], etc., where genotype to phenotype associations are maintained and regularly updated. There have been several new tools made available which can detect variations without the need for assembling the genome for the individual, such as those used in the 1000 Genome Project consortium which finds great applicability in case the coverage of sequences is low [1] and has, so to speak, yet have a profound impact at a population level. In this article, we share the results of the variations detected in a family of four individuals' viz., father, mother, and two daughters.

## 2    Methods

The blood samples of a family were collected in Amsterdam, although they might not be individuals who are of direct Dutch descent as Amsterdam is a cosmopolitan city. To keep their identities anonymous, we will refer to them as A105A, A105B, A105C, and A105D, respectively. The DNA was extracted and sequenced on Illumina HiSeq sequencer with an average coverage of more than 12x across the genome and with the raw read length of 90 bases at either end of the paired end reads with an average insert size of about 470 bases. As there are many copies of mitochondrial DNA in a cell, the sequencing coverage of mitochondrial DNA would be several folds higher than 12x. The reads were then assembled into respective contigs using parallel assembler ABySS [43] version 1.3.1 with optimal parameters of kmer size (k) of 49 and minimum reads to make a con-sensus contigs (n) of 3 to yield highest possible N50 value for the contigs ~3000. Default values of SOAPdenovo [44] were also used for assembling the genomes. SSPACE[42] was also used for assembly. On average it required about 140 GB of RAM in a shared environment and 49 computing wall-clock hours on a symmetric multiprocessor cluster with 6 computing cores each of capacity 2.6 GHz. The assemblies of the four individuals were then aligned globally to the NCBI human reference genome, Build 37, followed by extraction of SVs information of category insertions and deletions only (InDels), and single nucleotide polymorphisms (SNPs) on regions of misalignment [20,21]. Genome plot for the A105 A, B, C & D is shown in Figure A where one can graphically obtain estimates of regions of alignments and misalignments with the reference genome NCBI HuRef build 37.

The total time for the alignment and extraction of information on a single computing core of 2.6 GHz capacity came out to about 85 wall-clock hours, for each assembly.

## 3    Results

### 3.1    Scientific Outcomes

The SNPs of the whole genome can also be used for generic ancestry and divergence determination. For this reason, we also conducted a SNP genome-wide inheritance analysis and found that child A105C & child A105D had about 20.82% & 21.02%, respectively, of its SNPs that were novel compared to those present in the parents i.e. A105A & A105B (R code for analysis provided in supplementary). The consistency in both the children of about 21% SNPs being different from those found in either of the parents gives confidence in the methodology deployed. The difference could also be attributed to the aspect of only 14x genome average coverage and thus less robustness in detecting sequencing errors. Below in Figure B is the plot of the sum of the bases of InDels and SNPs for chromosome 20 of the A105 family. The sum of the bases of InDels have increased in the children when compared to their parents while the levels of SNPs remain more-or-less the same, clearly speaking of higher variance and thus higher standard deviation from mean in the InDels compared to the SNPs. This interesting observation of higher standard deviation in the InDels compared to SNPs, even in just one generation, clearly points out that InDels would be a stronger candidate for attributing a genetic signature of an individual and can be thus linked more confidently with determining the sex of an individual and associating with traits and disease phenotypes.

Figure C is a plot of the sum of the bases of SVs and SNPs respectively for the whole genome determined for the A105 family. The corresponding values for the plot along with mean and standard deviation values for SVs and SNPs for the family is shown in Table D., here again, we notice that the children have a relatively higher number of bases for SVs than their parents, though the levels of SNPs remain more-or-less the same. This finding thus proposes that even in one generation of the offspring, there can be a significant rearrangement in the genetic background to produce greater variance (and thus standard deviation) in the variation of genotype and thereby having an effect on phenotypes, and that the children are not an exact clone of the set of chromosomes they inherit from either parent as there will be significant variation, even when simply compared to the chromosomes of the parents that they inherited. The changes in SNPs are more restricted than insertions or deletions, and thus SVs serve as a stronger mean as a fingerprint and characteristics of an individual when analyzed genome wide.

In Figure D and Figure E you will see that the calls of bases on the Y chromosome of InDels and SNPs, respectively, is far higher for the father than the mother or the two daughters, thereby clearly being able to differentiate male from female. The pseudoautosomal regions, PAR1, PAR2, are homologous sequences of nucleotides on the X and Y chromosomes [28] where Genetic recombination (occurring during sexual reproduction) is known to be limited only to the pseudoautosomal regions (PAR1 and PAR2) of the X and Y chromosomes. Thus, small matches will be expected to be found in the female candidates when making calls for SVs and SNPs having the reference Y-chromosome from NCBI Build 37 as the counter alignment pair, as observed in our data and plotted in Figures D & E. It is also observed that the difference in the calls of sums of bases for InDels is far higher than the calls for the sum of the bases for SNPs, thereby proposing that the former is a stronger means to determine the sex of an individual than the latter. This also proposes that contrary to what is observed genome-wide, the SVs have higher selection pressure than the SNPs in the Y-chromosome. This strong selection pressure of Y-chromosome associated SVs compared to SNPs in woman, as also shown in Table A where the SVs comprise only about 0.44% of total SVs in male while the SNPs comprise about 23% of that number of male, which would usually be those of PAR1 and PAR2 region, proposes a stronger method by SVs to determine paternal inheritance in a woman, by the data illustrated here.

Even though this work involved had about 14x on average coverage for the genome, since the mitochondrial DNA comprises anywhere from 2 to 10 copies of the DNA for each mitochondrion [38], and since each cell in itself comprises of 1000s of mitochondria, the effective coverage for a mitochondrial genome would be in the order of 14x * Mean(2,10) * 1000, which is Order(14x*6*1000), which is about Order(84,000x). With such a heavy coverage of the mtDNA, it is highly unlikely that the assembly process of the mtDNA genome would be faulty given that most assembly work typically ranges for genome coverage in the median value of 49x with a population size of 27 as per [39]. From the already existing knowledge of inheritance of mitochondrial DNA, one would expect all the SNPs and SVs successful calls in mother to be found in all the children as well, as mitochondrial DNA is known to be maternally inherited. This is because mitochondrial

DNA material is present in the cytosol of a cell and not in the nucleus, and there is a lesser possibility for the cytosol of the sperm cells to integrate with the cytosol of the mother ova and is known to be destroyed at fertilization. So, for determining maternal inheritance, mtDNA is the same as his mother's mtDNA, which is the same as her mother's mtDNA and so on.

Our findings for A105 family analysis revealed contradicting results. Not all SNPs and SVs present in the mother were found to be present in the children. There were cases found where a SNP was found to be present in the father and a child but not in mother. Table B shows the list of SNPs and Table C shows the list of SVs in A105 family. This proposes a discovery that mitochondrial DNA can have paternal sources of inheritance as well, though they can also be a result of de-novo genetic changes rather than inheritance. Further, comparing Table B and C, it is discovered by the observation that mitochondrial DNA is less prone to SVs than SNPs, and that can be possibly attributed to the fact that mitochondrial DNA is not exposed to the phenomenon of crossing-over of genetic material as is the case with chromosomes. This proposes a discovery that mitochondrial DNA can have paternal or mixed paternal-maternal sources of inheritance as well, though most likely they can be a result of de-novo genetic changes rather than inheritance. If de-novo mutations happen to be the case, it would again oppose the 'mtDNA bottleneck' theory which states that mtDNA is subjected to high bottleneck [29]. The results by Table B & C would then propose a 'mtDNA bottleneck leakage' theory, where possibly one of the factors that were proposed in [29] has caused a bottleneck for mutation that must have gone uncontrolled. The heteroplasmy nature of mtDNA cannot be ruled out either, but the fact that maternal mtDNA SNPs are different than each of the offsprings raises several questions. Further, the ratio of SVs bases calls to the size of genome is significantly less for mitochondrial DNA ( of the order of $2.35*10^{-4}$ ) than for the whole genome ( of the order of $2.5*10^{-3}$), thereby providing further evidence that structural variations in mitochondrial DNA has higher selection pressure than the rest of the genome and is thus a more rare event in the mitochondria relative to the rest of the genome. This ratio remains comparable to the rest of the genome when considered for SNPs (of the order of $5.3*10^{-4}$ for mitochondria and the order of $8.0*10^{-4}$ for whole-genome).

Mitochondrial DNA and Y-chromosome DNA have been widely used to determine maternal and paternal ancestry respectively, such as in recent findings for Native Americans and Indigenous Altaians [22]. Based on the discoveries above, it can thus be safely concluded that if we continue with ancestry determination by mitochondrial DNA, then SVs would serve as better means to determine ancestry for a longer period than SNPs, as they are relatively more rare events. At the same time, the SNPs of the mitochondria would serve as a better candidate for the characteristic signature of the individual and can be used to determine ancestry and divergence for a relatively shorter period. Having said that, it would still be pro-posed that given that there is the possibility of mitochondrial DNA to be inherited by the father as well; maternal ancestry determination by mtDNA should be rephrased as simply ancestry determination by mtDNA. This will also mean that all the analysis, which different scientists across the globe have been conducting so far assuming mtDNA to be maternally inherited, will need a complete change in the understanding and knowledge generated. As it is confirmed that the Y-chromosome is completely paternally inherited, ancestry determination by 'Y line tests' as Y-chromosomes are confirmed to be inherited from the father remain a good methodology. Further, as observed and stated above, since SVs have higher selection pressure than SNPs for the Y-chromosome, the SVs will serve as a better means for paternal ancestry determination for a relatively longer time-span and the SNPs would serve as a better candidate to determine paternal ancestry and divergence in a relatively shorter time-span.

## 3.2 Clinical Applications

If the immunologic responses after the grafting of an organ from a donor to the receptor are known before conducting the transplant, we can be more predictive of the chances of success of the transplantation. The immunologic responses are dictated by the MHC region of the genome, which in humans corresponds to the HLA domain in chromosome VI.

DNA editing technologies such CRISPR/Cas9 [41], in the hope that one day the method would be even more precise, combined with this kind of study, as mentioned in this paper for detecting any unwanted structural variations, can be a great promise for the future. For this purpose, we first matched the SNPs from the A105 family members to the known OMIM[30] database and stored the result in Microsoft Excel sheet as shown in Figure G. From this table, filtering out those entries that have clinical association is straightforward, as shown in Figure H. Databases such as dbSNP[32] version 130 was also used for the very purpose and to ClinVar[31, 37] database version 20140211. Figures I, J, K & L show the results of ClinVar analysis for A105 A, B, C, & D having 26, 25, 31 & 24 entries respectively. From this data, we could see that apart from SNPs of clinical significance that was inherited from the parents, both children A105C & A105D also had 7 and 5 novel clinically relevant SNPs due to de-novo mutation (calculation in R script in supplementary). Figures M & N show the novel SNPs.

## 4   Discussions

The motivation of the current work is to lay the foundation for a personalized medicine scientific age, demonstrating clearly that the nuts-and-bolts needed for personalized diagnostics is well in place for us to confidently enter the commercialization stage of the technology. Although 14× coverage is far higher than the previous studies of 1000 genome project phase 1, which had lower coverage, fu-ture similar studies can be conducted with coverage of a recommendation of 50×
if genome assembly approach before analytics is to be deployed. The results of
mtDNA analysis seemed to not be affected by these factors as 1000s of copies of
mtDNA exist in the cell anyways. The clinical results generated by the analysis of
non-mitochondrial DNA cannot be trusted with complete sanctity for coverage of
only 14x, and so those medical outcomes were not discussed here. Certainly, the
work should not be left to be purely automated and the results generated by the
methods described here should also be validated by a team of experts. The limita-
tion of the work also questions the need for cheaper computing facilities with
more high-performance memory for the purpose, especially for genome assembly
work. Techniques, such as mapping, read directly to the reference genome can al-
so be used as an alternative, which does not require a high amount of memory, but
those techniques come with their own set of problems such as missing out any
novel region and their genomic loci information in the test genome, which might
be missing in the reference genome.

## 5    Conclusion

This research article improves our understanding of human genetics, variations
in the genome, and inheritance. 'Mitochondrial Adam' theory can well co-exist
with 'Mitochondrial Eve' theory as proposed by Lewin et. al. [48] given that we
now have evidences of paternal mtDNA inheritance, while we can continue with
'Y-chromosomal Adam' theory given that both the SVs and SNPs sum of nucleo-
tide variance and standard deviation point out to the same findings. We also pro-
posed that given that the standard deviation of the sum of nucleotides in SVs are
larger than that of SNPs, SVs differences would better characterize the gender of
an individual. It provides us with new scopes to fetch relevant information and
opens the door for many newer technologies to be built based on the discoveries.
The discoveries make us more equipped with statistical and robust, efficient and
relatively less costly means to derive information such as sex determination, or
immunologic response to disease, or success rate of organ transplant, or suscepti-
bility to diseases and possible cure for them. Given that there is a larger standard
deviation of the sum of the nucleotides of InDels (SVs) than sum of SNPs, this can
greatly and strongly impact how personalized genomic analysis and diagnostics
are being carried out, as the scientific community develops algorithms and tools to
better utilize this knowledge shared by this paper.

## 6    Abbreviations

SVs - Structural Variations, InDels – Insertions and Deletions, DIPs – Deletion
and Insertion Polymorphism, SNPs - Single Nucleotide Variations, mtDNA - mi-
tochondrial Deoxyribonucleic acid, CNV - Copy Number Variations, SNVs - Sin-
gle Nucleotide Variations

# 7    Declaration Section

**Ethics approval and consent to participate**

The names of individuals participating in the research were kept anonymous, and their consent was taken for analysis work to be published. No medical outcome of the work was reported. This paper did not need any additional ethics committee approval written or verbal, since this is an extended version of papers [23, 24, 49], and the same data was used, for more detailed interpretation of results which has been reported in this paper.

**Consent for publication**

Consent to publication was taken at the time when bio-samples were being collected. This paper did not need any additional consent for publication written or verbal, since this is an extended version of papers [23, 24, 49], and the same data was used, for more detailed interpretation of results which has been reported in this paper.

**Availability of data and Supplementary materials**

All supporting data are provided in this article. Supplementary materials can be downloaded from https://sites.google.com/a/iitdalumni.com/abi/educational-papers .

**Competing interests**

None to be declared.

**Funding**

This paper did not need any additional data acquisition since this is an extended version of papers [23, 24, 49], and the same data was used, for more detailed interpretation of results, and thus no additional funding was needed.

**Authors' contributions**

All work, idea generation, coding, analysis of results and writing the paper has been done by the first author.

**Acknowledgements**

Author is thankful to Ms Kelin Coleman for proof-checking the manuscript before submission.

## 8    References

1. 1000 Genomes Project Consortium et al. A map of human genome variation from population scale sequencing. Nature 467, 1061-1073 (2010).

2. Mills, R.E. et al. Mapping copy number variation by population scale sequencing. Nature published online, doi:1:10.1038/nature09708 (3 February 2011).

3. Fanciulli, M. et al. FCGR3B copy number variation is associated with susceptibility fo systemic, but not organ-specific, autoimmunity. Nat. Genet. 39, 721-823 (2007).

4. Aitman, T.J. et al. Copy number polymorphism in Fcgr3 predisposes to glomerulonephritis in rats and humans. Nature 439, 851-855 (2006).

5. Gonzalez, E. et al. The influence of CCL3L1 gene-containing segmental duplications on HIV-1/AIDS susceptibility. Science 307, 1434-1440 (2005).

6. Fellermann, K. et al. A chromosome 8 gene-cluster polymorphism with low human beta-defensin 2 gene copy number predisposes to Crohn disease of the colon. Am. J. Hum. Genet. 79, 439-448 (2006).

7. Yang, Y. et al. Gene copy-number variation and associated polymorphisms of complement component C4 in human systemic lupus erythematosus (SLE): low copy number is a risk factor for  and high copy number is a protective factor against SLE susceptibility in European Americans. Am. J. Hum. Genet. 80, 1037-1054 (2007).

8. Hollox, E.J. et al. Psoriasis is associated with increased beta-defensin genomic copy number. Nat. Genet. 40, 23-25 (2008).

9. Feuk L, Carson AR, Scherer SW: Structural variation in the human genome. Nat Rev Genet 2006, 7:85-97.

10. Bodmer W, Bonilla C: Common and rare variants in multifactorial susceptibility to common diseases. Nat Genet 2008, 40:695-701.

11. Iafrate AJ, Feuk L, Rivera MN, Listewnik ML, Donahoe PK, Qi Y, Scherer SW, Lee C: Detection of large-scale variation in the human genome. Nat Genet 2004, 36:949-951.

12. Sebat J, Lakshmi B, Troge J, Alexander J, Young J, Lundin P, Maner S, Massa H, Walker M, Chi M, Navin N, Lucito R, Healy J, Hicks J, Ye K, Reiner A, Gilliam TC, Trask B, Patterson N, Zetterberg A, Wigler M: Large-scale copy number polymorphism in the human genome. Science 2004, 305:525-528.

13. Redon R, Ishikawa S, Firch KR, Feuk L, Perry GH, Andrews TD, Fiegler H, Shapero MH, Carson AR, Chen W, Cho EK, Dallaire S, Freeman JL, Gonzalez JR, Gratacos M, Huang J, Kalaitzopoulos D, Komura D, MacDonald JR, Marshall CR, Mei R, Montgomery L, Nishimura K, Okamura K, Shen F, Somerville MJ, Tchinda J, Valsesia A, Woodwark C, Yang F, et al.: Global variation in copy number in the human genome Nature 2006, 444:444-454.

14. Tuzun E, Sharp AJ, Bailey JA, Kaul R, Morrison VA, Pertz LM, Haugen E, Hayden H, Albertson D, Pinkel D, Olson MV, Eichler EE: Fine-scale structural variation of the human genome. Nature 2006, 444:444-454.

15. Khaja R, Zhang J, MacDonald JR, He Y, Josheph-George AM, Wei J, Rafiq MA, Qian C, Shago M, Pantano L, Aburatani H, Jones K, Redon R, Hurles M, Armengol L, Estivill X, Mural RJ, Lee c, Scherer SW, Feuk L: Genome assembly comparison identifies structural variants in the human genome. Nat Genet 2006, 38:1413-1418.

16. Korbel JO, Urban AE, Affourtiti JP, Godwin B, Grubert F, Simons JF, Kim PM, Palejev D, Carriero NJ, Du L, Taillon BE, Chen Z, Tanzer A, Saunders AC, Chi J, Yang F, Carter NP, Hurles ME, Weissman SM, Harkins TT, Gerstein MB, Egholm M, Snyder M: Paired-end mapping reveals extensive structural variation in the human genome. Nat Genet 2006, 38:1413-1418.

17. Kidd JM, Cooper GM, Donahue WF, Hayden HS, Sampas N, Graves T, Hansen N, Trague B, Alkan C, Antonacci F, Haugen E, Zerr T, Yamada NA, Tsang P, Newman TL, Tuzun E, Cheng Z, Ebling HM, Tusneem N, David R, Cillett W, Phelps KA, Weaver M, Saranga D, Brand A, Tao W, Gustafson E, McKernan K, Chen L, Malig M, et al.: Mapping and sequencing of structural variation from eight human genomes. Nature 2008, 453:56-64.

18. Conrad DF, Pinto D, Redon R, Feuk L, Gokcumen O, Zhang Y, Aerts J, Andrews TD, Barnes C, Campbell P, Fitzgerald T, HuM, Ihm CH, Kristiansson K, Macarthur DG, Macdonald JR, Onyiah I, Pang AW, Robson S, Stirrups K, Valsesia A, Walter K, Wei J, Tyler-Smith C, Carter NP, Lee C, Scherer SW, Hurles

ME: Origins and functional impact of copy number variation in the human genome. Nature 2010, 464:704-712.

19. Buchanan JA, Scherer SW: Contemplating effects of genomic structural variation. Genet Med 2008, 10:639-647.

20. Abhishek Narain Singh, Comparison of Structural Variation between Build 36 Reference Genome and Celera R27c Genome using GenomeBreak, Poster Presentation, The 2nd Symposium on Systems Genetics, Groningen, 29-30 September 2011

21. Abhishek Singh, GENOMEBREAK: A versatile computational tool for genome-wide rapid investigation, exploring the human genome, a step towards personalized genomic medicine, Poster 70, Human Genome Meeting 2011, Dubai, March 2011

22. Dulik MC, Zhadanov SI, Osipova LP, Askapuli A, Gau L, Gokcumen O, Rubinstein S, Schurr TG, Mitochondrial DNA and Y Chromosome Variation Provides Evidence for a Recent Common Ancestry between Native Americans and Indigenous Altaians, Am J Hum Genet. 2012 Feb 10;90(2):229-46. Epub 2012 Jan 25.

23. Abhishek Narain Singh, A105 Family Decoded: Discovery of Genome-Wide Fingerprints for Personalized Genomic Medicine, page 115-126, Proceedings of the International Congress on Personalized Medicine UPCP 2012 (February 2-5, 2012, Florence, Italy), Medimond Publisher, ScienceMED journal vol.3 issue 2, April 2012.

24. Abhishek Narain Singh, Customized Biomedical Informatics, BIOCOMP'13, The 14th International Conference on Bioinformatics and Computational Biology, July 22-25, 2013, Las Vegas, USA http://tinyurl.com/BIOCOMB13 .

25. Wilson, Ian J. et al. "Mitochondrial DNA Sequence Characteristics Modulate the Size of the Genetic Bottleneck." Human Molecular Genetics 25.5 (2016): 1031–1041. PMC. Web. 28 Dec. 2017.

26. Sallevelt SCEH, de Die-Smulders CEM, Hendrickx ATM, et al
De novo mtDNA point mutations are common and have a low recurrence risk
Journal of Medical Genetics Published Online First: 22 July 2016. doi: 10.1136/jmedgenet-2016-103876

27. Li M et. al, Transmission of human mtDNA heteroplasmy in the Genome of the Netherlands families: support for a variable-size bottleneck. Genome Res 2016;26:417–26. doi:10.1101/gr.203216.115

28. Mangs, Helena; Morris BJ (2007). "The Human Pseudoautosomal Region (PAR): Origin, Function and Future". Current Genomics. 8 (2): 129–136. doi:10.2174/138920207780368141

29. Khrapko K. Two ways to make a mtDNA bottleneck. Nature genetics. 2008;40(2):134-135. doi:10.1038/ng0208-134.

30. Amberger J, Bocchini CA, Scott AF, Hamosh A (2009) McKusick's Online Mendelian Inheritance in Man (OMIM). Nucleic Acids Res 37: D793–6. Pmid:18842627

31. Landrum MJ, Lee JM, Riley GR, Jang W, Rubinstein WS, Church DM, et al. (2014) ClinVar: public archive of relationships among sequence variation and human phenotype. Nucleic Acids Res 42: D980–5. Pmid:24234437

32. Sherry ST, Ward MH, Kholodov M, Baker J, Phan L, Smigielski EM, et al. (2001) dbSNP: the NCBI database of genetic variation. Nucleic Acids Res 29: 308–311. Pmid:11125122

33. Whirl-Carrillo M, McDonagh EM, Hebert JM, Gong L, Sangkuhl K, Thorn CF, et al. (2012) Pharmacogenomics knowledge for personalized medicine. Clin Pharmacol Ther 92: 414–417. Pmid:22992668

34. Thorisson GA, Lancaster O, Free RC, Hastings RK, Sarmah P, Dash D, et al. (2009) HGVbaseG2P: a central genetic association database. Nucleic Acids Res 37: D797–802. Pmid:18948288

35. The UniProt Consortium (2014) Activities at the Universal Protein Resource (UniProt). Nucleic Acids Res 42:D191–D198. Pmid:24253303

36. "A map of human genome variation from population-scale sequencing" Nature 467 1061-1073 2010

37. Landrum MJ, Lee JM, Benson M, Brown G, Chao C, Chitipiralla S, Gu B, Hart J, Hoffman D, Hoover J, Jang W, Katz K, Ovetsky M, Riley G, Sethi A, Tully R, Villamarin-Salomon R, Rubinstein W, Maglott DR. ClinVar: public archive of interpretations of clinically relevant variants. Nucleic Acids Res. 2016 Jan 4;44(D1):D862-8. doi: 10.1093/nar/gkv1222. PubMed PMID:26582918

38. Wiesner et al. Biochem Biophys Res Commun 1992 , 183 (2):553-9

39. EKBLOM, R., WOLF, J.B.W. A field guide to whole-genome sequencing, assembly and annotation. Evolutionary Applications, v. 7, n. 9, p. 1026-1042, 2014.

40. Abhishek Narain Singh, Variations in Genome Architecture, Poster, International Congress on Personalized Medicine, 2-5 Feb UPCP 2012, Florence, Italy

41. Horvath P, Barrangou R (January 2010). "CRISPR/Cas, the immune system of bacteria and archaea". Science. 327 (5962): 167–70.

42. Marten Boetzer, Christiaan V. Henkel, Hans J. Jansen, Derek Butler, Walter Pirovano, Scaffolding pre-assembled contigs using SSPACE, *Bioinformatics*, Volume 27, Issue 4, 15 February 2011, Pages 578–579, https://doi.org/10.1093/bioinformatics/btq683

43. Simpson, J. T., Wong, K., Jackman, S. D., Schein, J. E., Jones, S. J., & Birol, I. (2009). ABySS: a parallel assembler for short read sequence data. *Genome research*, *19*(6), 1117–1123. doi:10.1101/gr.089532.108

44. Luo, R., Liu, B., Xie, Y., Li, Z., Huang, W., Yuan, J., … Wang, J. (2012). SOAPdenovo2: an empirically improved memory-efficient short-read de novo assembler. *GigaScience*, *1*(1), 18. doi:10.1186/2047-217X-1-18

45. Abhishek Narain Singh, Precision Genomics and Biomedical Discoveries, PAPER ID #: BIC4132, The 20th International Conference on Bioinformatics and Computational Biology, July 29- August 1, 2019, Las Vegas, USA

**46. S. Luo et al., "Biparental inheritance of mitochondrial DNA in umans," PNAS, https://doi.org/10.1073/pnas.1810946115, 2018. https://www.pnas.org/content/115/51/13039**

47. Schwartz M, Vissing J. Paternal inheritance of mitochondrial DNA. N Engl J Med. 2002;347:576–80.

48. Lewin R (1987), "The unmasking of mitochondrial Eve", *Science*, **238** (4823): 24–26, doi:10.1126/science.3116666, PMID 3116666

49. Abhishek Narain Singh, Customized Biomedical Informatics, Springer Nature, BMC, Big Data Analytics, May 2018 https://bdataanalytics.biomedcentral.com/articles/10.1186/s41044-018-0030-3

50. Abhishek Narain Singh, Precision Genomics and Biomedical Discoveries, PAPER ID #: BIC4132, The 20th International Conference on Bioinformatics and Computational Biology, July 29- August 1, 2019, Las Vegas, USA